\documentclass[final,5p,twocolumn]{elsarticle}

\usepackage{amssymb}

\journal{International Journal of Hydrogen Energy}

\begin{document}

\begin{frontmatter}



\title{The Design of Long-Life, High-Efficiency PEM Fuel Cell Power Supplies for Low Power Sensor Networks}


\author{Jekanthan Thangavelautham$^1$, Daniel Dewitt Strawser$^2$, Steven Dubowsky$^2$}

\address{$^1$School of Earth and Space Exploration, Arizona State University, AZ, 85287 USA e-mail:jekan@asu.edu; \\ $^2$Mechanical Engineering Dept., Massachusetts Institute of Technology, MA, 02139.}

\begin{abstract}
Field sensor networks have important applications in environmental monitoring, wildlife preservation, in disaster monitoring and in border security.  The reduced cost of electronics, sensors and actuators make it possible to deploy hundreds if not thousands of these sensor modules.  However, power technology has not kept pace.  Current power supply technologies such as batteries limit many applications due to their low specific energy. Photovoltaics typically requires large bulky panels and is dependent on varying solar insolation and therefore requires backup power sources.  Polymer Electrolyte Membrane (PEM) fuel cells are a promising alternative, because they are clean, quiet and operate at high efficiencies.  However, challenges remain in achieving long lives due to factors such as degradation and hydrogen storage.  In this work, we devise a framework for designing fuel cells power supplies for field sensor networks.  This design framework utilize lithium hydride hydrogen storage technology that offers high energy density of up to 5,000 Wh/kg.  Using this design framework, we identify operating conditions to maximize the life of the power supply, meet the required power output and minimize fuel consumption.  We devise a series of controllers to achieve this capability and demonstrate it using a bench-top experiment that  operated for 5,000 hours.  The laboratory experiments point towards a pathway to demonstrate these fuel cell power supplies in the field.  Our studies show that the proposed PEM fuel cell hybrid system fueled using lithium hydride  offers at least a 3 fold reduction in mass compared to state-of-the-art batteries and 3-5 fold reduction in mass compared to current fuel cell technologies.

\end{abstract}


\end{frontmatter}


\section{Introduction}
\label{intro}

\begin{figure} [h!]
\centering
\includegraphics[width=3.5in]{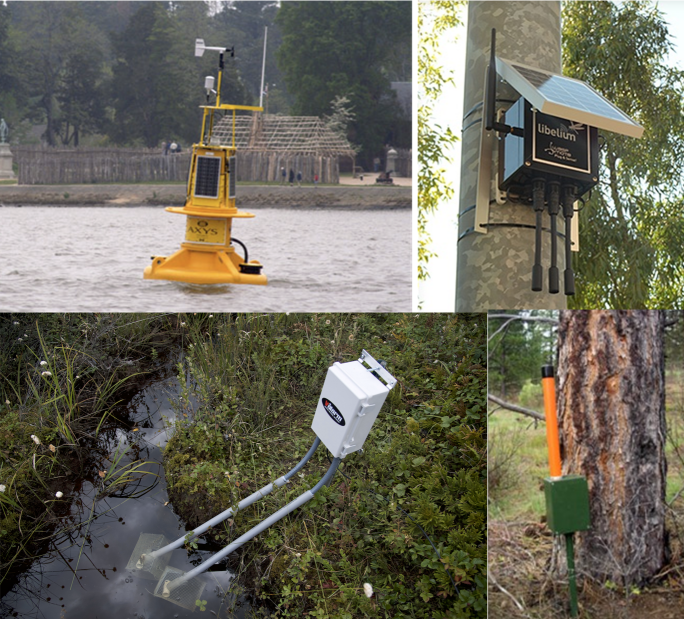}
\caption{Sensor networks have important applications in environment monitoring including monitoring of water, air and soil health.}
 \label{fig:sensor_net}
\end{figure}

Ubiquitous sensor networks operating in unstructured field environments have many important applications including exploration and mapping of inaccessible environments, environmental monitoring including measuring air, water and soil quality, disaster prediction of forest fires, avalanches, earthquakes and volcanoes and in wildlife monitoring to protect endangered species (Figure~\ref{fig:sensor_net}) \cite{aky}.  These sensor networks could be used to improve agricultural production, by closely monitoring soil humidity, crop health and making predictions for harvest. In the security sphere, sensor networks maybe used for monitoring borders for illegal entry or smuggling of dangerous goods.  Several hundreds or thousands of modules might be deployed over large areas and wirelessly report to a base-station.  To be practical, the systems need to be low-cost, reliable and operate unattended, ideally for years. Significant advances have been made in wireless communication, miniaturization of electronics, autonomous control systems, sensors and actuators.  However, advancement in power supplies have not kept pace.

Conventional batteries do not meet the needs of many types of wireless sensor network due to their low specific energy.  For long life missions, they need to be recharged or replaced.  While significant work is being done to address their energy limitations \cite{ritchie,tarascon}, rechargeable batteries alone cannot meet the high energy requirements of  long duration field sensors.
Photovoltaics is another important power source but these devices are bulky, require periodic maintenance such as cleaning of panels and requires a steady source of solar insolation throughout the year.  These factors limit their applications to certain climate zones or applications that can afford hibernation due to periods of inclement weather or winter. Typically, due to daily variabilities in solar insolation, photovoltaic systems need to be coupled with a battery.  Energy harvesting using piezoelectric generators has been proposed for field sensors, but have yet to be practically applied.  These generators require a steady source of vibrations and are suitable only for specific very low-power applications.  A better solution is needed.

Here, fuel cells are proposed as power source for wireless field sensor networks.  Fuel cell are electrochemical energy conversion devices that convert chemical energy directly into electricity \cite{barbir,hayre}.  Three types of fuel cells  are  attractive for sensor networks, they include PEM fuel cells, direct-methanol fuel cells and microbial fuel cells.  These fuel cells operate at low temperatures and are quiet.  Out of the three, PEM fuel cells are highly efficient operating at 60-65\%, react with hydrogen and oxygen to produce electricity. However, the storage and release of hydrogen is typically a challenge. In this work, we address this major challenge and propose a feasible solution.  Direct methanol fuel cells use methanol and oxygen from air to produce electricity.  They tend to have lower efficiencies than PEM, longer startup times and produce carbon dioxide and water.  Direct methanol fuel cells are attractive, because methanol is easy to store and has higher energy densities than conventional hydrogen storage.  The main challenge with conventional direct methanol fuel cells is an inherent limitation with the fuel cell design, compared to PEM.  The buildup of CO$_2$ accelerates degradation of the fuel cell and causes high rates of cross-over resulting in lower operating efficiencies and unreliability.

However, PEM fuel cells are not widely used in field applications because they also face significant challenges.  Firstly, PEM fuel cells are faced with the problem of degradation of their components that result in shortened lives and unreliability compared to batteries.  A second major challenge is the storage of hydrogen~\cite{schlap}.  Conventional methods of hydrogen storage are bulky and inefficient, providing only a marginal advantage over current batteries.   A third major challenge is that these fuel cells produce lower power compared to batteries.  A fourth challenge is that PEM fuel cells have high costs.  Significant progress is being made in all these areas. Our research addresses the first three challenges.

In this paper, we present PEM fuel cells as a promising solution to powering sensor networks for long duration.  The PEM fuel cell power supply is implemented as a fuel cell hybrid system, held under controlled conditions to maximize life, maximize cell operating efficiency and minimize component degradation.  This method enables the fuel cell power supply to achieve conversion efficiencies of 60-65 \% and operating lives of 3-5 years.  The fuel cell is supplied with hydrogen fuel from a water activated lithium hydride hydrogen generator that freely extracts water vapor from the air.  This method offers a theoretical specific energy of 5,000 Wh/kg, nearly 40 times that of conventional lithium ion batteries. Our case studies presented in this paper show that through effective design and control that the proposed fuel cell power supply can be superior to conventional batteries for field sensor network applications.

The remainder of this article is organized as follows.  Section 2 presents background and related work. Section 3 presents power management, air management and fuel management design and control of the PEM fuel cell power supplies to achieve long life. Section 4 presents an experimental system used to test the concept.  Section 5 presents several case studies and discussion comparing the concept to conventional fuel cells technologies and batteries and Section 6 presents our conclusions.

\section{Background}
\label{background}

Fuel Cells have been proposed as power supplies for field sensor networks. An important factor in their selection is that they are clean, highly efficient, offer high specific energy and are quiet~\cite{barbir,hayre}.  These factors make them well suited for deployment in sensitive, pristine environments.  In contrast, conventional primary batteries pose concerns of leakage into the environment.  As a result, fuel cells are an interesting choice for environmental sensing.  Proposed technologies to power sensor networks in the field include Polymer Electrolyte Membrane (PEM) fuel cells, Direct Methanol Fuel Cells (DMFC) and Microbial Fuel Cells~\cite{logan,piet,vaghari,barbir,hayre}.

PEM have been proposed as low-power sources for use in the field \cite{chraim,devaraj,thanga,strawser}.  Impressively, MEMs scale PEM fuel cells have been fabricated and tested for this purpose \cite{zhi}.  These systems integrate several components of a fuel cell power supply into a chip sized wafer, including the fuel and oxygen source.  These fuel cells have been shown to operate for a few hundred hours, though long-life experiments have yet to be tested.  Further, these systems tradeoff small size for performance, including power output and current density. Control of these fuel cells is performed using passive systems that self-regulate fuel dispensation and oxygen.  They typically lack a control system to maximize the operating life of the fuel cell.

Work by Chraim and Karaki proposes PEMs to power wireless sensor networks \cite{chraim}.  One of the major challenges in the implementation of PEM fuel cells for field applications is the storage of hydrogen.  Several works have proposed use of metal hydrides from cannisters.   One approach involves utilizing ambient power sources to electrolyze water into hydrogen and oxygen \cite{devaraj}.  The stored hydrogen and oxygen is fed to a fuel cell to enable power generation during off-hours~\cite{devaraj}.  Another proposed application~\cite{magno} demonstrates a hybrid system where a fuel cell recharges a battery that is in turn used to power a sensor network.  The fuel cell keeps the battery topped up.  Our work identifies some of the additional benefits of this technology particularly to avoid oscillations in fuel cell voltage that limits life~\cite{thanga}.

Microbial Fuel Cells~\cite{logan} use microbes and their metabolism to generate electricity.  They are well suited for marsh, ponds and  even reservoirs containing waste water.  Although the power density tends to be low,  the system benefits from its relative simplicity~\cite{logan}.  A key challenge, though, is the maintaining the well-being of the microbial organisms to ensure sufficient power is generated.  Death of a microbial colony due to disease or change in environmental conditions can have an impact on the microbial fuel cell and hence this requires human oversight/tending~\cite{logan}.

\begin{figure} [h!]
\centering
\includegraphics[width=3.5in]{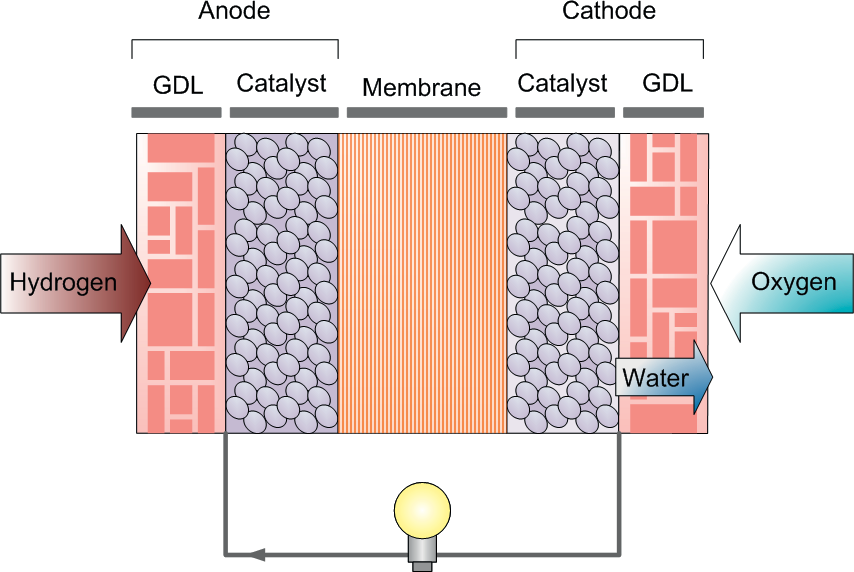}
\caption{A Polymer Electrolyte Membrane (PEM) fuel cell and its major components.}
 \label{fig:fc_components}
\end{figure}

Our focus has been on PEM fuel cells because it is one of the more well-studied and mature technologies that shows important potential for both high-energy and long-life applications ~\cite{hayre,barbir}.  However, degradation of PEM fuel cells is a challenge.  A PEM fuel cell has several major components that are all subject to degradation (Figure~\ref{fig:fc_components}).  They are the Gas Diffusion Layer (GDL), bi-polar plates, the membrane and catalyst layers.  The GDL facilitates transfer of input gasses to the anode and cathode.  The bi-polar plates have an important role in distributing the reactant gases to the anode and cathode, conduct electrical current from the cell and help to remove heat from the active area, while preventing leakage of gasses~\cite{hermann}.  The anode catalyst layer facilitates the oxidation of hydrogen molecules into protons while the membrane allows for the transport of protons from the anode to the cathode. The cathode catalyst layer facilitates the assembly of protons and oxygen molecules into water via a reduction reaction.

Extensive research has been done to identify the mechanisms that degrade fuel cell components.  GDL degradation affects the ability of the cell to absorb reactants.  This degradation includes loss of hydrophobicity of the cathode that causes flooding~\cite{frisk} and loss in fuel cell power performance.  Degradation of the GDL can reduce or block gas passageways resulting in choking of the fuel cell.  A major source of GDL degradation has been due to mechanical compressions resulting in stress and strains that reduce the micropore regions~\cite{kandlikar}.  This reduces gas transport and thus reduces the effectiveness of the GDL.  Freeze thaw cycles are well known to damage the GDL as freezing water expands and damages the micro-pores.  Another source of damage to the GDL is the loss of Polytetrafluoroethylen (PTFE)~\cite{kandlikar,hayre}.  PTFE is used to maintain gas passages by repelling water using its hydrophobic properties.  The PTFE is weakly bonded to carbon fibre material and therefore high temperatures and corrosion of the carbon layers due to oxidation can slowly erode them.

The erosion of the PTFE results in accumulated flooding in the fuel cell~\cite{kandlikar}.  Flooding is known to cause corrosion of various components of the fuel cell including the catalyst~\cite{madden}. Structural damage, such as from freezing~\cite{hou}, mechanical stress, wear and tear of a GDL can reduce the cell's ability to absorb fuel or oxidizer~\cite{wu} and thus reducing a cell's performance and resulting in catalyst degradation.  However, it should be noted that some of these sources of GDL degradation can be prevented by avoiding such conditions as structural damage or freezing.

Bi-polar plates are also known to corrode.  They are known to corrode and produce secondary effects that may exacerbate other forms of degradation in the fuel cell and in power generation.  Rates of bi-polar plate corrosion are governed by the pH of the water and this may change with degradation of membrane that results in higher acidity of the water~\cite{houb}.  Build up of oxide films on the bi-polar plates can result in increased resistance which results in lower power output from the fuel cell~\cite{papa}.  Further, this can result in the leaching of ion species that can interact with other corrosion processes.

Membranes are subject to degradation and can be classified into three categories, thermal, mechanical and chemical~\cite{liu,wu,laconti,collier}.
An important source of membrane degradation is due to mechanical stress and strain.  Membranes once weak are prone to formation of pinholes that result in fuel cross-over and significant reduction in power generated.  This is followed by catastrophic failure of the fuel cell \cite{liu,huang}.  Mechanical stress and strain is known to occur due primarily to humidity and temperature cycling \cite{huang}.  Humidity is known to impact the mechanical properties of the membrane. With too high a humidity, the membrane curls up and with too little humidity, the membrane hardens and drys out.   The impact of relative humidity cycling can be severe.  It has been shown that a membrane cycled between a Relative Humidity of 30 \% and 80 \% faces structural failure after only a 100 cycles~\cite{huang}.  A second source of membrane degradation is due to chemical attack \cite{liu}.  The end result from chemical attack is reduced strength of the membrane leading to structural failure.

In addition, membranes can degrade due to migration of impurities, particularly catalyst particles that deposit in the membrane.  This can result in local structural weak points that cause pinholes, permitting cross-over of reactant gases and their direct combustion of the reactants causing loss of power, rapid increase in temperature and fuel cell death~\cite{huang}.  The root cause of the above failure is the migration of carbon and catalyst particles due to catalyst degradation.  Hence in many cases platinum catalyst degradation in the cathode is an important factor in fuel cell durability and life~\cite{wu}.  One cause of catalyst degradation is the dissolution of the platinum particles into ions~\cite{wu}.  The ions either redeposit on large platinum particles (similar to Ostwald ripening) or dissolve and migrate away from the catalyst layer and into nearby regions~\cite{shao}. Sustained degradation reduces the available catalyst surface in the anode and cathode resulting in loss of power.

Catalyst degradation in the cathode can be a major PEM degradation mechanism  because it can result in irreversible loss of power and cause structural failure of the membrane.  Extensive experimental evidence has shown that fuel cell catalyst degrades due to several operating factors including high operating voltages, high temperature, and high humidity \cite{rubio}. Physical models have been developed that describe dissolution of the catalyst, thought to be the primary cause of fuel cell degradation \cite{bi}.   In our earlier work, we focused on developing a physical model of fuel cell catalyst degradation that matches existing experimental data, enabling us to make long term predictions of life and performance under field conditions~\cite{bi,thanga}.  As discussed above, PEM fuel cells are delicate and require balancing various system parameters such as fuel flow rate, temperature, humidity and operating voltage within a narrow operational window.  Operation outside of this narrow window can result in irreversible damage to a fuel cell.

Our focus is to maximize both conversion efficiency and life.  This is critical for low power sensor network application, where the objective is to maximize energy density of the power source.  This requires the fuel cell to operate at higher, but constant voltages.

Current fuel cell control techniques are designed  for stationary, large-scale power generation applications and use conventional feedback control, gain scheduling and set-point-tracking approaches towards operating fuel cells~\cite{puk}. In these configurations, the fuel cell voltage may vary drastically depending on the power demand, which according to our models can substantially shorten life.  PEM-battery hybrid-systems are a promising solution to this problem.  In our proposed approach, the fuel cells are small, of low power and operate at constant operating voltage that constantly charges a battery that handling high and varying loads.  The battery protects the fuel cell from external oscillations in load that shorten life.  In addition, the battery provides burst of high power depending on load demand.  A constant operating point simplifies air, fuel and water management, instead of requiring active control methods as in~\cite{puk}.

A second major challenge as noted earlier is the problem of hydrogen storage.  High pressure and cryogenic storage of hydrogen are clearly impractical for small, low-power applications such as sensors networks.  A third option is the use of metal hydrides.  Conventional reversible metal hydrides release  hydrogen through changes in pressure or temperature.  An alternate option is the use of chemical hydrides that release hydrogen through chemical reaction~\cite{kong}. While reversible hydrides are
valued because of their ability to be recharged with hydrogen, they are not ideal for long-life
field applications because they normally have low hydrogen storage densities (defined as the
weight of the hydrogen divided by the total weight of the hydride) on the order of 1-2\% \cite{chandra}.  However there exists other higher hydrides with much higher yield but they require much higher temperatures, in the order 200 -700 $^{o}$ C.  Achieving such high temperatures for small low-power devices increases complexity.

\begin{figure} [h!]
\centering
\includegraphics[width=3.5in]{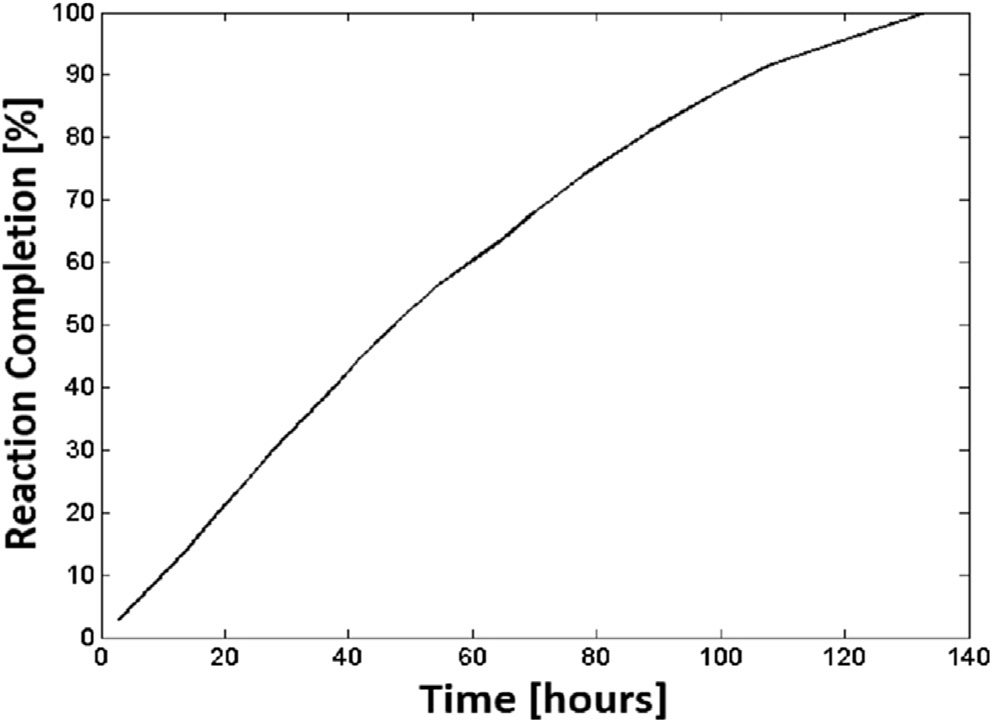}
\caption{Lithium hydride hydrolysis reaction with water.  The results show that lithium hydride can achieve reaction completion~\cite{strawser}. }
 \label{fig:reaction_completion}
\end{figure}

Hydrolysis is the reaction of chemical hydride with water to produce hydrogen ~\cite{strawser,kong}.  However, popular water activated metal hydrides including sodium borohydride (NaBH$_4$)~\cite{sch} and magnesium hydride (MgH$_2$)~\cite{kojima} have low hydrogen content, low reliability and require expensive catalysts.  Alternatives such as calcium hydride do not require catalyst but have low yield.  Our research has focused on Lithium Hydride (LiH) which has higher net hydrogen content by mass than calcium hydride.  Hydrogen can be released by exposing lithium hydride to water releasing the hydrogen from the hydride and stripping water of its hydrogen. Lithium hydride, unlike other water activated hydrides, requires no complex mechanisms or catalysts to start, control and complete the hydrogen release reaction~\cite{strawser,kong}.  Our experimental studies show that water activated lithium hydride can achieve 95-100 \% reaction completion rates with excess of water (Figure~\ref{fig:reaction_completion})~\cite{strawser}.

Together, these approaches point towards a pathway for developing long-life, high-efficiency, low-power fuel cell power supplies for sensor networks.  In the following section, we shall describe the challenges and implementation steps of the power, air, water and fuel management systems for the fuel cell power supply.

\section{Long Life, Low Power Fuel Cell Power Supply Design}




\subsection{Fuel Cell Degradation}

Fuel cell degradation is an important factor in the design of the fuel cell power supply and these are impacted by operating conditions, such as voltage, temperature, humidity and voltage oscillations ~\cite{thanga}.   Using the catalyst degradation model developed in~\cite{thanga}, we determine the effect of operating conditions on fuel cell catalyst life. The rate of change in surface area of the catalyst particles is used to estimate the catalyst life of the fuel cell. Here, the effective end of the fuel cell catalyst life is assumed to occur when the catalyst surface area has decreased by 25\%~\cite{thanga} which is a more stringent assumption than from the US Department of Energy protocols~\cite{doe}.  By the time, 25\% of the catalyst area has dissolved it is estimated that the secondary effects of catalyst dissolution is well under way causing eventual mechanical failure of the membrane.  The fuel cell life is taken to be the expected time before the fuel cell reaches the terminal phase of its life leading to catastrophic failure.

The model is run and we quantify the effect of each operating condition on fuel cell catalyst life, while keeping other conditions constant.  Use this model, we obtain life predictions for the fuel cell catalyst. Next, we fit an analytical expression to these life prediction curves and use the expression to extrapolate fuel cell catalyst life  under field conditions.

\begin{figure} [h!]
\centering
\includegraphics[width=3.5in]{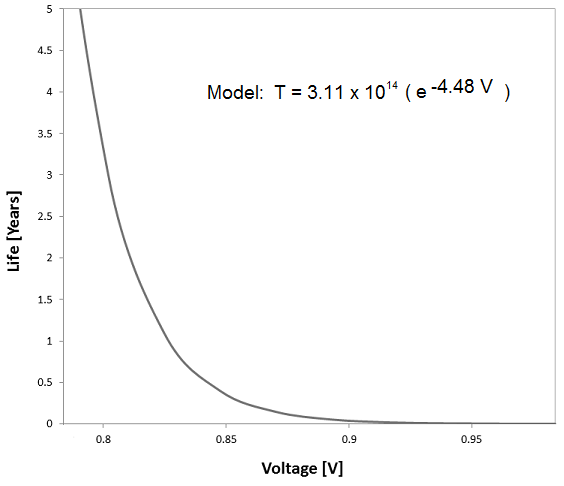}
\caption{The predicted effects of operating voltage on fuel cell catalyst life (cathode humidity 50\%, temperature 25 $^{o} C$.}
 \label{fig:voltage}
\end{figure}

Figure~\ref{fig:voltage} shows that fuel cell catalyst life exponentially increases for lower voltages as shown. To achieve 3 years of life, the fuel cells would  have to operate at 0.8 V or less.  Increased voltages as had been determined from the model accelerates dissolution of the platinum catalyst, thus reducing its electrochemically active surface area.  However operating at high voltages increases the fuel cell conversion efficiency, but also results in less power output.  These conflicting factors need to be considered to determine a suitable operating voltage.

\begin{figure} [h!]
\centering
\includegraphics[width=3.5in]{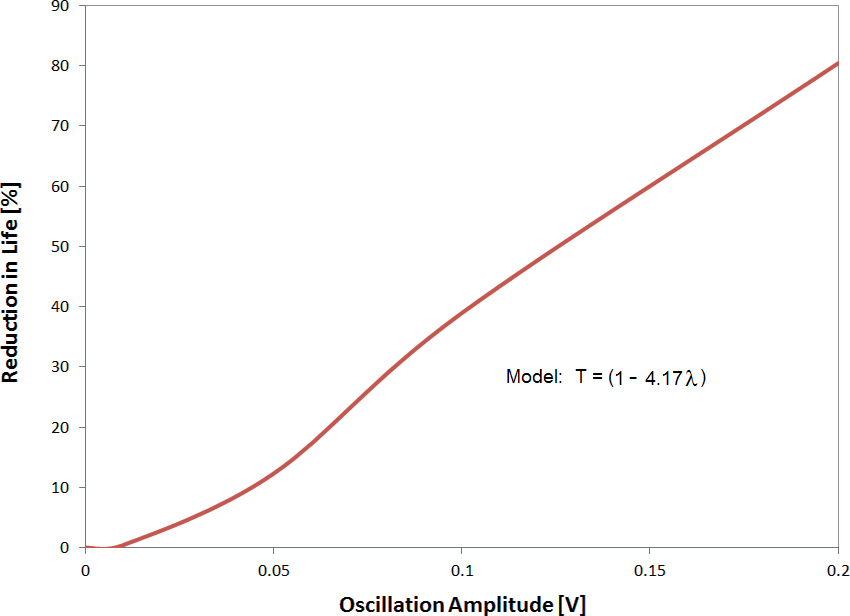}
\caption{The predicted effects of voltage oscillation on fuel cell life.}
 \label{fig:oscillation}
\end{figure}

Typically all but the simplest of electrical devices have varying electrical loads.  The work in~\cite{thanga} analyzed the effect of voltage fluctuations on fuel cell catalyst life.  The results show a linear reduction in catalyst life for a linear increase in voltage oscillation amplitude.  To understand the full impact of voltage oscillations, the percentage reduction in life value needs to be multiplied to the expected life in Figure~\ref{fig:oscillation} to obtain an absolute effect on catalyst life.

\begin{figure} [h!]
\centering
\includegraphics[width=3.5in]{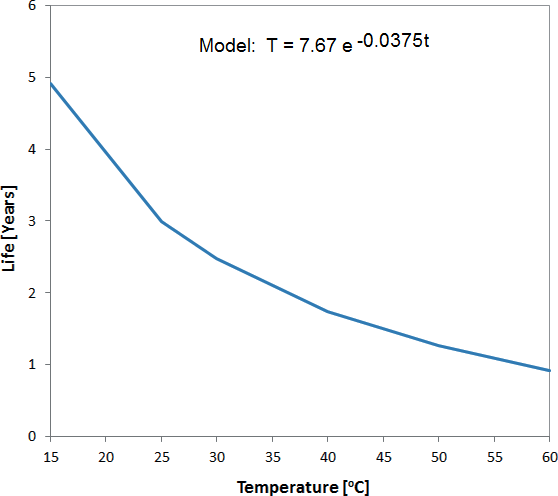}
\caption{The predicted effects of temperature on fuel cell life.}
 \label{fig:temperature}
\end{figure}

The effect of temperature on fuel cell performance is critical for field devices where temperatures will vary over the course of a day and over seasons.  The effect of temperature on fuel cell catalyst life is shown in Figure~\ref{fig:temperature}.  For increased temperature, an exponential decrease in life is predicted by the model.  For example, operating at 0.8 V the life is 4.5 times shorter operating at 60 $^{o}$ C than for 15 $^{o}$ C.  Low temperatures decreases catalyst degradation and increases life, but can produce other problems such as increased condensation resulting in flooding.

\begin{figure} [h!]
\centering
\includegraphics[width=3.5in]{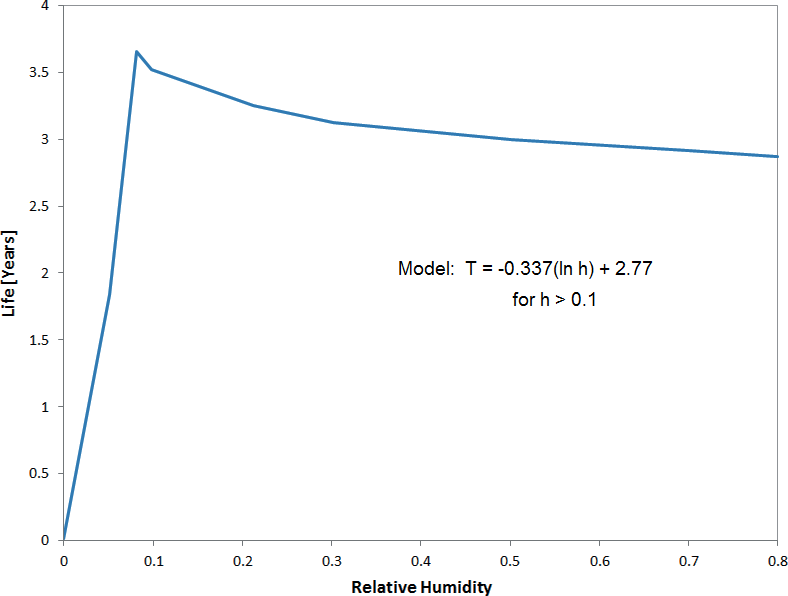}
\caption{The predicted effects of cathode humidity on fuel cell life.}
 \label{fig:humidity}
\end{figure}

Humidity is an important operating parameter for fuel cell catalyst life.  The effect of cathode humidity on the catalyst life of an air breathing PEM fuel cell is shown in Figure~\ref{fig:humidity}.  As seen, life is significantly shortened when the relative humidity approaches 0, while a peak occurs at 10 \% relative humidity.  Further increase in humidity is less substantial.
This result is caused by varying rate of platinum oxide formation due to cathode humidity.

Based on the degradation models and analysis of the individual operating conditions and effect on fuel cell, we produce a combined equation to predict the effect of fuel cell catalyst life.  This equation for $\tau_{clife}$ in years  presumes each operating variable has a independent effect on the life of the fuel cell catalyst.  The equation is given below:

\begin{equation}\label{eq:excavation_fitness}
\tau_{clife} = {a_V}{e^{ - {k_V}V}} \cdot {a_T}{e^{ - {k_T}T}} \cdot \left( {{a_h}\ln h + {k_h}} \right) \cdot (1 - {a_\lambda }\lambda )
\end{equation}

The constants for the equations are given in Table~\ref{tab:eq_variables}.
\begin{table}
\centering
\begin{center}
\begin{tabular}{ |l|l| }
  \hline
  Variable & Value \\
  \hline
  $a_V$ & $3.11 \times 10^{14}$ \\
  $a_T$ & 7.67 years \\
  $a_h$ & $-3.37 \times 10^{-1}$ \\
  $a_\lambda$ & $4.173\; V^{-1}$ \\
  $k_h$ & $2.77$ \\
  $k_V$ & $-4.48\; V^{-1}$ \\
  $k_T$ & $-3.75 \times 10^{-2}\; {^{o}C}^{-1}$ \\
  \hline
\end{tabular}
  \caption{Fuel cell catalyst life model parameters}
  \label{tab:eq_variables}
\end{center}
\end{table}

\begin{table*}[t]
\centering
\begin{center}
\begin{tabular}{ |l|c| }
  \hline
  Conditions & Predicted Fuel Cell Life \\
  \hline
  Operating Voltage: 0.78 V, 0.2 V oscillation, No Environment Control & 0.27 years  \\
  Operating Voltage: 0.78 V, 0.2 V oscillation, 15\% Humidity, +5 $^{o}$C Dew Point & 2.2 Years\\
  Operating Voltage: 0.78 V, 0.02 V oscillation, 15\% Humidity, +5 $^{o}$C Dew Point & 12.2 Years\\

  \hline
\end{tabular}
  \caption{Fuel cell catalyst life comparison for field sensor network}
  \label{tab:field_sensor_table}
  \end{center}
\end{table*}

The variable takes into account the effect of operating voltage, operating temperature, operating humidity and voltage oscillations. For field applications, ambient temperature and humidity will change over the course of a day and over the seasons.  A second degradation phenomena modelled is the degradation of the membrane due to humidity cycling and is given below.  Based on the experiments from~\cite{huang} and others,  the PEM fuel cell membrane can only withstand a finite number of humidity cycles that results in stress loading culminating in mechanical failure of the membrane.

\begin{equation}\label{eq:membrane_life}
\tau_{mlife} = \Delta t_{hosc}\cdot \frac{b_{memb}}{RH_{max}-RH_{min}}
\end{equation}

where $\Delta t_{hosc}$ is humidity cycling period, $b_{memb}$ is a membrane specific constant and is 40 for Nafion NR111~\cite{huang}, $RH_{max}$ and $RH_{min}$ is the maximum and minimum relative humidity. The total life of the fuel cell is then modelled as the following:

\begin{equation}\label{eq:total_life}
\tau_{life} = min(\tau_{clife}, \tau_{mlife})
\end{equation}

where $\tau_{life}$ is the expected life of the PEM fuel cell.  In other words, the expected life is the minimum of the expected catalyst life presented earlier and expected life of the membrane due to humidity cycling.  These factors independently impact the fuel cell.   When the change in humidity is minimized, the number of humidity cycles  before membrane failure increases until it is not the life limiting factor for the fuel cell.  Catalyst degradation begins with performance degradation of the catalyst and finally result in catastrophic loss, while membrane degradation results in mechanical damage to the membrane that results in the formation of pinholes and ends up in catastrophic loss.

\begin{figure} [h!]
\centering
\includegraphics[width=3.5in]{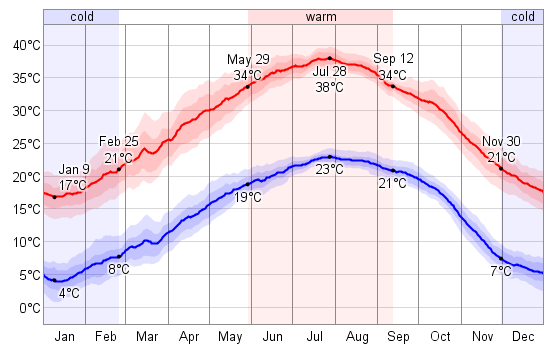}
\caption{Annual daily maximum and minimum temperature in Negev, Israel}
 \label{fig:negev_temp}
\end{figure}

\begin{figure} [h!]
\centering
\includegraphics[width=3.5in]{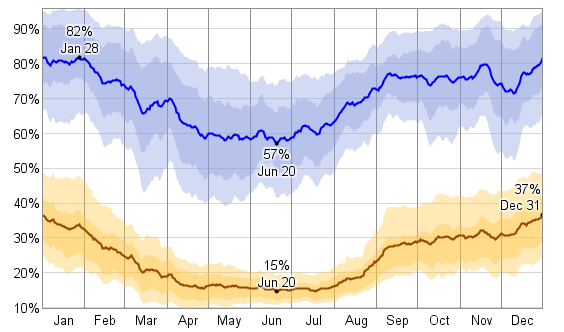}
\caption{Annual daily maximum and minimum humidity in Negev, Israel}
 \label{fig:negev_humid}
\end{figure}

Figure~\ref{fig:negev_temp} shows the maximum and minimum daily temperatures expected throughout the year in Negev, Israel.  Figure~\ref{fig:negev_humid} shows the maximum and minimum daily humidity expected in Negev.  Using these varying temperature and humidity conditions, we can predict the net effect on life of a fuel cell power supply in the field.  We presume the fuel cells operating at 0.78 V and voltage oscillations are at 0.2 V.  Using ambient temperature and humidity (no environment control), the expected fuel cell catalyst life is 1.4 years (Table~\ref{tab:field_sensor_table}).  Further, if we account for daily humidity cycling on membrane life, that life drops to 0.27 years which is 2,400 hours and so this becomes the life limiting factor.

\begin{figure} [h!]
\centering
\includegraphics[width=3.5in]{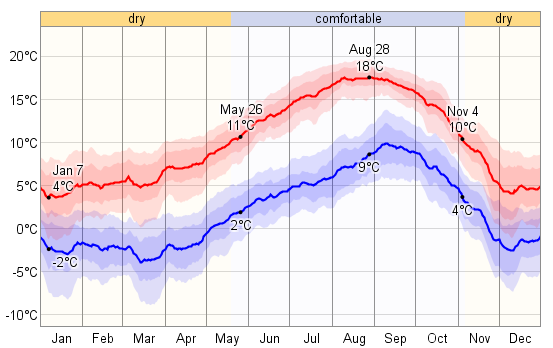}
\caption{Annual daily dew point temperature in Negev, Israel}
 \label{fig:negev_dew}
\end{figure}

Next, we set the fuel cell to operate at a setpoint of 15 $\%$ relative humidity and operate the power supply 5 degrees above dew point.  The annual maximum and minimum dew point in Negev, Israel is shown in Figure~\ref{fig:negev_dew}.  By effectively lowering the operating temperature and humidity, we expect to increase the life of the fuel cell.  In addition, lower operating humidity also reduces the chances of flooding.  The resultant fuel cell life is 2.2 years (Table~\ref{tab:field_sensor_table}).

Further we can reduce voltage oscillations experienced by  the fuel cell system to 0.02 V.  In Section 4 we show our hybrid controller to achieve maximum oscillations of 0.02 V and this results in a further extension in life to 12.2 years (Table~\ref{tab:field_sensor_table}).  In these later two examples, we have eliminated humidity cycling and its impact on the membrane by maintaining the fuel cell at constant humidity.   By effectively controlling the humidity, temperature, operating voltage and voltage oscillations, the fuel cell power supply is predicted to operate for long lives.  Under such conditions, other life-limiting factors will impact the life of the fuel cell and these include the slow but gradual degradation of the PTFE in the GDL ~\cite{kandlikar}.

\begin{figure} [h!]
\centering
\includegraphics[width=3.5in]{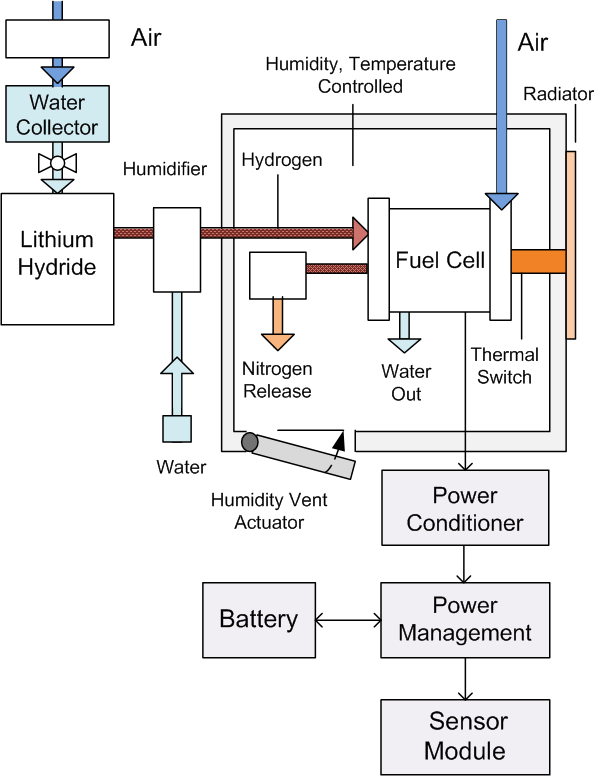}
\caption{Proposed fuel cell power supply.}
 \label{fig:fuel_cell_benchtop_block_diagram}
\end{figure}

In the following section, we  describe how we implement a fuel cell control system to achieve long life.  A block diagram of the fuel cell power supply is shown in Figure~\ref{fig:fuel_cell_benchtop_block_diagram}.  The fuel cell power supply consist of several important components, a hydrogen generator,  fuel cells, humidity and temperature controller, air  management system, and  power management controller.  In the following section, the power management system will be analyzed.

\subsection{Power Management}

The proposed fuel cell power supply consists of a fuel cell-battery hybrid system.  The fuel cell constantly charges a battery and the battery/fuel cell system periodically powers an electrical load.  By having several fuel cell in series, the net voltage can be set high enough to charge a battery without the use of additional electronics.  Otherwise, a step-up DC-DC convertor is required.  The DC-DC typically introduces voltage oscillations, that  by our analysis result in degradation of the fuel cell power supply~\cite{thanga}.  It also decreases the efficiency of the system due to voltage conversion loses.

Without the DC-DC convertor, the system is simpler and more efficient.  In addition, a variable resistor  is included in the circuit.  This variable resistor circuit is activated and  used during startup and shutdown, to ensure the fuel cell is at proper voltages to avoid fuel starvation.  The variable resistor ensures the fuel cell maintains a constant voltage during startup and shutdown.  Finally, an electrical load is connected to the circuit and periodically turned on at a set duty cycle or on demand.

First, the average power required of the fuel cell needs to be determined. Second, an operating voltage range has to be selected for the cells.  For sensor network applications, we wish to maximize both the life of the fuel cell and fuel cell conversion efficiency so as to minimize hydrogen fuel consumption.  The operating efficiency of a fuel cell is given by~\cite{barbir, hayre}:

\begin{equation}\label{eq:excavation_fitness}
\lambda_{FC} = 0.81 V
\end{equation}

where $V$ is the fuel cell cell operating voltage.

A third constraint is that the fuel cells need to have matching output voltage to charge the battery. This is to avoid a specialized battery charging circuitry.  A battery charging circuitry would add additional complexity to the system, and be a source of voltage oscillations and it will further reduce system efficiency due to conversion losses.  However, the tradeoff is that the fuel cells need to  be in a matching series configuration, with a high enough voltage to charge the battery.

A battery needs to be selected that has enough capacity so that when the load is powered, it does not result in a substantial voltage drain from the battery.  This is once again to avoid voltage oscillations that degrade the fuel cell.  This expected change in voltage due to loss of charge can be calculated from integrating a battery-voltage charge curve.

A power control system monitors the fuel cell output voltage and the battery (see Figure~\ref{fig:fuel_cell_voltage}).  Once the voltage drops just below the operating voltage range, the control system would divert output power from the fuel cell to charge the battery until the net voltage reaches the upper boundary of the operating voltage range.

\begin{figure} [h!]
\centering
\includegraphics[width=3.5in]{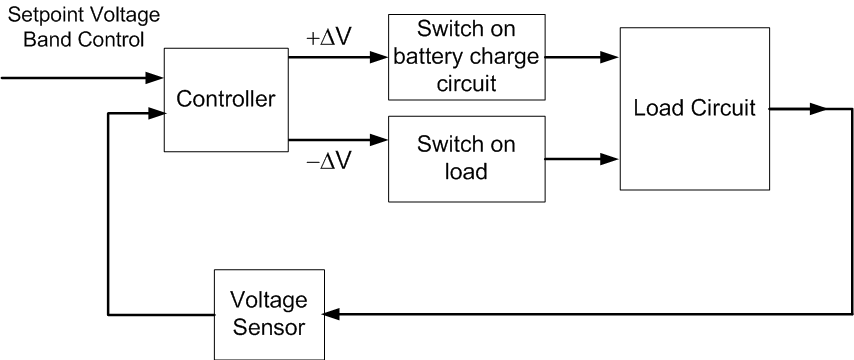}
\caption{Proposed fuel cell voltage control system.}
 \label{fig:fuel_cell_voltage}
\end{figure}

\subsection{Air and Water Management}

The purpose of the air management system is to ensure oxygen is delivered to the fuel cell cathode and ensure inert gasses such as nitrogen don't buildup in the anode.  We assume air-breathing PEM fuel cells are used, hence the oxygen is freely extracted from the air.  However, it is critical that the air entering the cathode maintains a proper humidity to ensure smooth operation of the fuel cell.  This requires that the air humidity  be not too low or not too high.  The effect of humidity on catalyst degradation from~\cite{thanga} is shown here.  The results show that the humidity needs to be above 10\% to avoid accelerated degradation.  However the humidity also cannot be too high, otherwise this might cause flooding that blocks pores in the GDL and result in reduced power output due to fuel starvation, oxygen starvation or both. For the micro-fuel cells considered here, this maximum humidity cannot exceed 70\%.  It is found that most of the waste water from the micro fuel cell exits the cathode.  Therefore this waste water needs to be disposed to prevent the cathode from reaching the maximum humidity.

A second objective as noted earlier is to prevent buildup of inert gasses such as nitrogen in the anode.
The fuel cells, as noted earlier, are configured in a dead-end anode mode.  This configuration typically maximizes fuel utilization, but at a cost of buildup of nitrogen at the anode.  If this is left uncontrolled, the fuel cells will starve of hydrogen and drop in voltage.  The net effect is that this degrades the fuel cell catalyst and limits life.  A conventional method of removing the nitrogen is using a purge valve, that is actively controlled.  The valve periodically opens and closes to dispose of the nitrogen according to a predetermined schedule \cite{puk}.  However, this requires active electronics and valves and results in some loss of hydrogen, up to 15 \%.  For low power systems, avoiding the active electronics and valves can both simplify the system and increase overall efficiency.

In our work, we focused on designing a passive membrane to continually purge nitrogen.  This is done using sufficiently thick foam, which lets nitrogen through but also some hydrogen.  The thickness of the foam is varied empirically  to let enough nitrogen out.  This is done by varying the foam and thickness and deterging voltage drops in dead mode of a test fuel cell.  In addition, the foam membrane is positioned at the dead-end, where nitrogen concentrates, while hydrogen is consumed by the fuel cells.  Through this design process, our work suggests, that nitrogen can be effectively purged with estimated hydrogen leakage losses of  5\%.

\subsection{Thermal Management}

\begin{figure} [h!]
\centering
\includegraphics[width=3.5in]{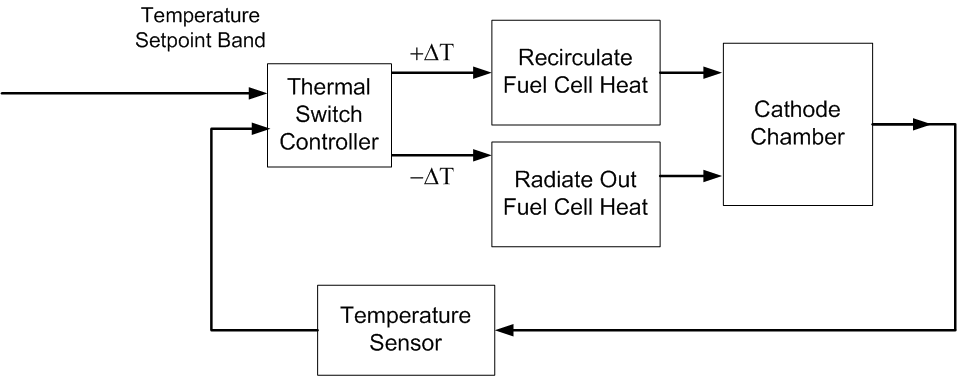}
\caption{Proposed fuel cell thermal control system.}
 \label{fig:fuel_cell_temp}
\end{figure}

As noted earlier, maintaining the fuel cell at a controlled temperature can substantially decrease fuel cell degradation.  It avoids build up of hot-spots, drying and reduction in humidity.  In our studies, we presume the fuel cell power supply is located in conditions where the temperature is mostly above freezing.  The proposed thermal controller is shown in Figure~\ref{fig:fuel_cell_temp}.  In this controls approach, a thermal switch is introduced.  The system works by operating within a temperature set-point band.  In an active setup, a temperature sensor is used to take readings and feed to a thermal switch, while in a passive setup, thermal expansion is used to close the switch.  When there is too much heat, the thermal switch is closed and is used to radiate the heat to an external radiator.  When the switch is open, internal heat generated from the fuel cell is recirculated.

\subsection{Humidity Management}

\begin{figure} [h!]
\centering
\includegraphics[width=3.5in]{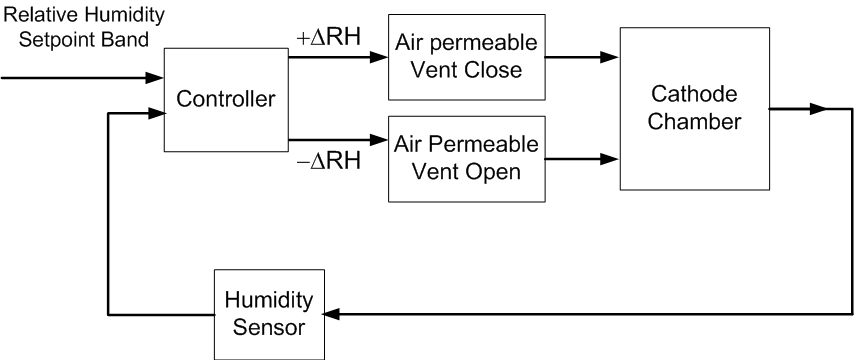}
\caption{Proposed fuel cell humidity control system.}
 \label{fig:fuel_cell_humid}
\end{figure}

As noted earlier, maintaining the fuel cell at a suitable humidity can decrease fuel cell degradation. Humidity needs to be maintained at a setpoint to minimize humidity cycling, which decrease membrane life.  However, the humidity needs to be high enough to ensure the membrane is conductive, yet not too high to avoid flooding. Flooding of the cathodes can disrupt fuel cell operations or cause long-term structural damage.  In our studies, we presume the fuel cell power supply is situated above freezing temperatures.  The proposed humidity controller is shown in Figure~\ref{fig:fuel_cell_humid}.  In this controls approach, a humidity controller constantly monitors the cathode humidity using humidity sensors.  Once the humidity reaches beyond a setpoint threshold, a vent opens to a dryer environment where the humidity is dissipated.  This control approach, just like the thermal control system, can be made passive utilizing humidity sensitive material that can expand or contract depending on humidity to open or close a vent.  However such methods are unable to achieve humidity control within +/- 10\% relative humidity.

\subsection{Fuel Management}

Here it is shown that lithium hydride is ideal for storage and release of hydrogen.  Hydrogen can be released by exposing the hydride to water releasing the hydrogen from the hydride and stripping water of its hydrogen according to the following reaction:

\begin{equation}\label{eq:lih}
{\rm{LiH}} + {\rm{H}}{}_2{\rm{O}} \to {\rm{LiOH}} + {{\rm{H}}_2}
\end{equation}

Lithium hydride unlike other water activated hydrides requires no complex mechanisms or catalysts to start, control and complete the reaction~\cite{strawser,kong}.  Our experimental studies show that water activated lithium hydride can achieve 100 \% reaction completion rates (see Figure~\ref{fig:reaction_completion}).  Another appealing feature of water activated lithium hydride for PEM fuel cells is that the fuel cell produces enough waste water for activating the lithium hydride.  When exhaust water from a fuel cell is reused for producing more hydrogen using a lithium hydride generator, the reaction achieves a theoretical 25 \% hydrogen storage efficiency or 5,000 Wh/kg energy density~\cite{strawser,Thangav2012} (40 folds higher than lithium ion batteries).

Based on the experiments performed, a semi-empirical model is developed describing the hydrolysis of thick layers of lithium hydride for the purpose of design of lithium hydride hydrogen generator~\cite{strawser,Thangav2012}.  The model predicts the total volume of hydrogen produced at a given time for a given humidity, volume, and exposed surface area of lithium hydride.  Using this model, we have developed a lithium hydride hydrogen generator design that can achieve reaction completion.

\subsection{Control of Hydrogen Generator}

\begin{figure} [h!]
\centering
\includegraphics[width=3.5in]{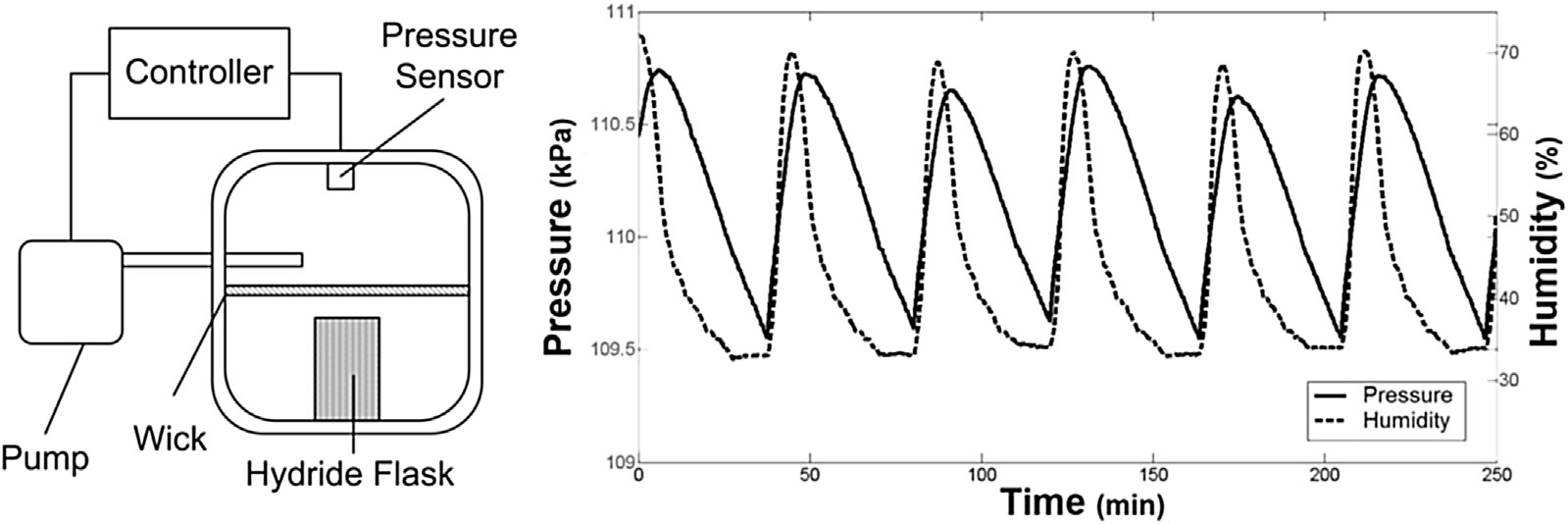}
\caption{Active control system used to control hydrogen generation and hydrogen pressure at 1.1 Bar}
 \label{fig:hydrogen_active}
\end{figure}

Several control strategies have been developed to produce the required hydrogen at high operating efficiencies.  Active control strategies had been initially pursued to achieve a desired hydrogen pressure.  A small peristaltic pump drawing an average power of 10$^{-2}$ mW is used to periodically dispense droplets of water exposed to the hydride to produce hydrogen for a 50 mW system (Figure~\ref{fig:hydrogen_active}). However, for low-power sensor network applications there is a need to simplify the system and increase its reliability by minimizing control electronics and actuators. Our work focuses on passive lithium hydride hydrogen generators that are simpler and more appropriate for low-power because they don't have active control components that otherwise require electrical power.

\begin{figure*} [t!]
\centering
\includegraphics[width=6.5in]{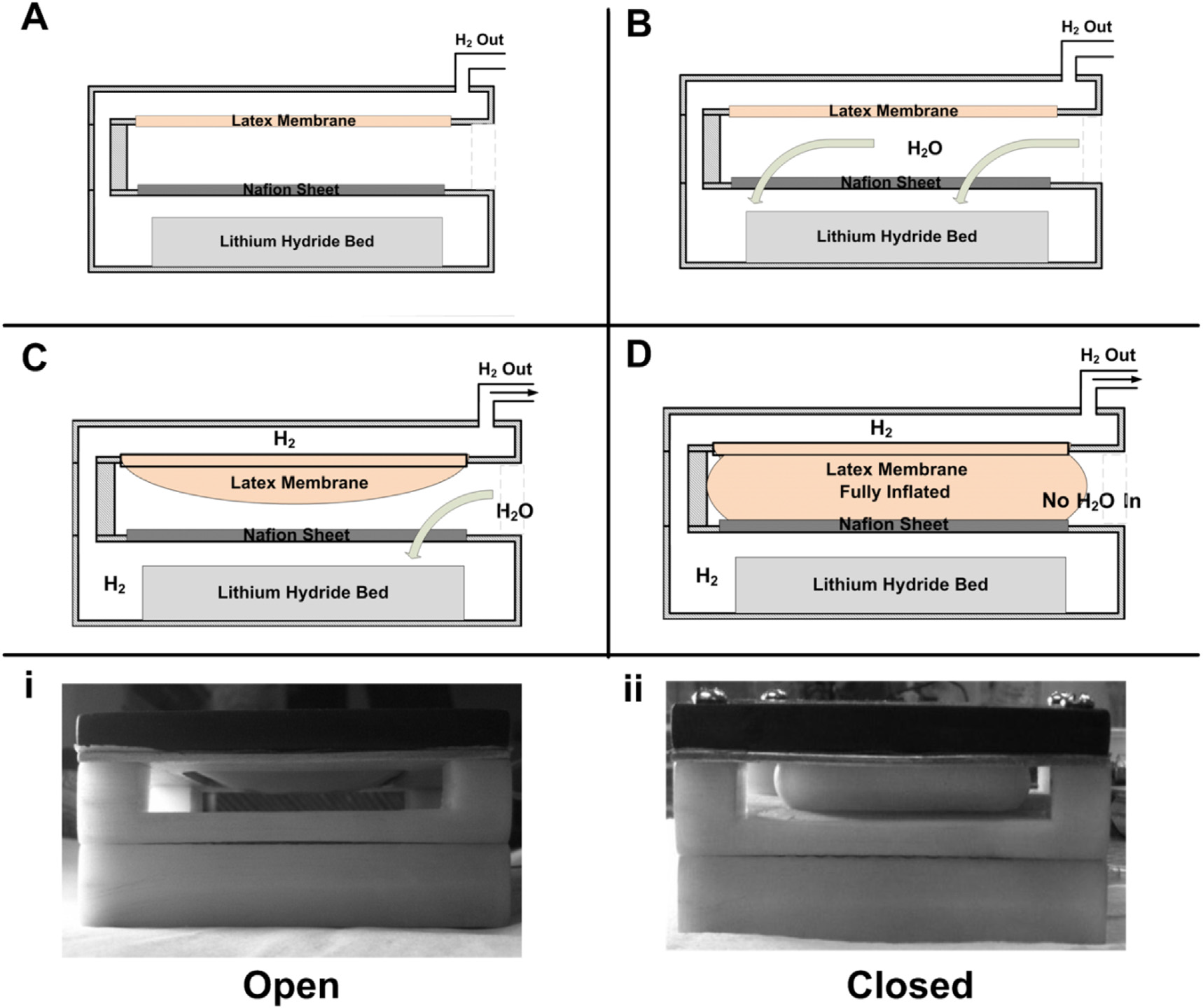}
\caption{Passive lithium hydride hydrogen generator design.}
 \label{fig:hydrogen_generator}
\end{figure*}

A passive control system works by exploiting the mechanics or dynamics of a material to perform control (Figure~\ref{fig:hydrogen_generator}). The proposed concept is to control the output hydrogen pressure.  The mechanism works by letting in water vapor through  Nafion$^{TM}$ membrane layer that produces hydrogen, which increases hydrogen pressure. The Nafion$^{TM}$ lets water vapor cross from a point of high vapor pressure to a point of lower partial pressure, but prevents the escape of hydrogen.  The partial pressure of water vapor  inside the chamber is low because it readily reacts with lithium hydride to produce hydrogen.  This setup in effect allows the lithium hydride to passively extract water vapor from the surrounding. Once the unit reaches the target pressure,  a latex diaphragm expands and seals the Nafion$^{TM}$ preventing  further transport of water vapor.

Cross-section view of a passive lithium hydride generator is shown in Figure~\ref{fig:hydrogen_generator}.  The figure shows the following:  A)  Generator consisting of two opposing compartments that allow for pressure communication between them.  B)  Water vapor, either produced at the fuel cell's cathode or from the environment, enters through the Nafion$^{TM}$ sheet.  C.)  The water vapor reacts with the lithium hydride.  The generated pressure travels to the upper chamber and inflates the latex membrane.  D.)  The rate limiting mechanism occurs when the latex membrane presses against the Nafion$^{TM}$, greatly reducing the amount of water vapor entering the system. (i) Latex membrane fully contracted, allowing water vapor to enter the system. (ii)  Latex membrane fully expanded at target pressure, not allowing water vapor to enter the system.

The passive hydrogen generator is designed as to not carry water on board; instead it extracts water vapor from its surroundings.  This design is tested for long duration to produce hydrogen for fuel cell power supply in Section 4.  While passive lithium hydride generators relying on liquid water have been developed, this generator is unique in its ability to use water vapor from the atmosphere or fuel cell exhaust~\cite{strawser,Thangav2012}.  Additionally, lithium hydride generators have not been experimentally validated for long periods of time or with a hybrid PEM fuel cell system.  Based on the lessons learned, an experimental system was built to demonstrate the fuel cell power supply for field sensor networks.

\section{Experimental System}

\begin{figure} [h!]
\centering
\includegraphics[width=3.5in]{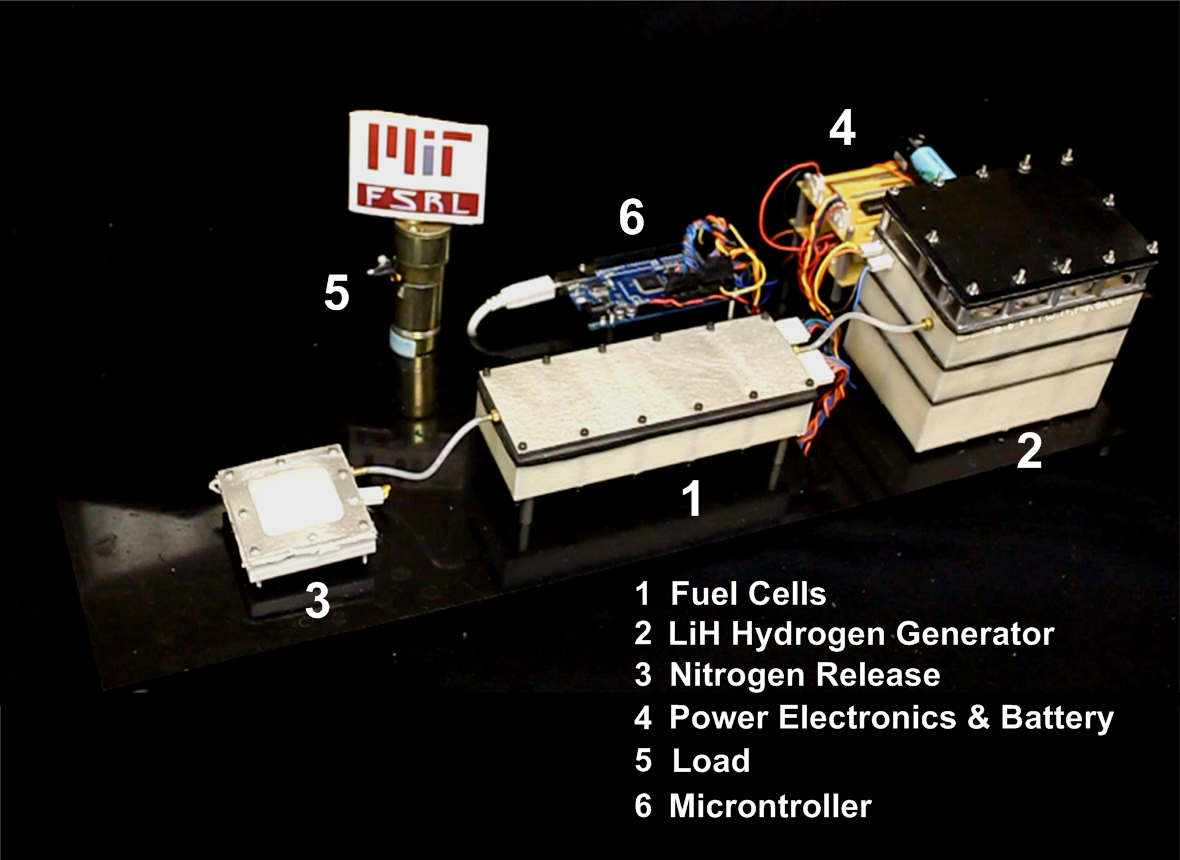}
\caption{Fuel cell power supply bench-top experiment}
 \label{fig:fuel_cell_benchtop}
\end{figure}

The fuel cell power supply concept presented in the previous section was tested using a revised bench top fuel cell system shown in Figure~\ref{fig:fuel_cell_benchtop}.  This revised benchtop system is being operated for 5,000 hours, much higher than our our previous experiments lasting 1,400 hours.  The system consists of (1) 5, 0.25 W Horizon Fuel Cells connected in series, a lithium hydride hydrogen generator (2), a passive nitrogen purge systems (3), fuel cell hybrid configuration containing an electronically programmable variable resistor and lithium ion rechargeable battery (4), a load consisting of a motor (5) and a micro-controller, containing fuel cell control algorithm and system to monitor and record experiment data from the humidity, voltage, pressure and temperature sensors.  The system is nominally designed to produce 250 mW.  The purpose of the experiment is to validate the fuel cell control system and hydrogen generator.

The system was designed to simulate the power supply for a remote sensor.  The sensor operates on a 10\% duty cycle, during which it switches on a motor.  This can be easily substituted to various environment sensor packages that require 0.25 to 0.5 W or less.

\subsection{Hydrogen Generator}

The hydrogen generator concept discussed above is pictured in Figure~\ref{fig:fuel_cell_benchtop}.  As stated above, optimization of system volume was not an objective of the experiment.  The generator's size was dictated by a two factors.  First, the exposed surface area of Nafion$^{TM}$ was designed to be twice as large as required to maintain the minimum flow rates.  The reason for this was to ensure that the generator produced sufficient hydrogen in a wide range of relative humidity (the room in which this experiment took place experienced 35-40 \% relative humidity).  Secondly, the hydrogen generator must be considerably taller than the height of the unreacted lithium hydride.  This is because the lithium hydride tends to expand 3 times its original volume when converted to lithium hydroxide.

\subsection{Fuel Cells}

Five fuel cells obtained from Horizon Fuel Cell Technologies, each with an MEA area of 5 cm$^2$ and max power output of 250 mW.  The fuel cells are connected in series and total system voltage is maintained at 3.75-3.8 V, which is 0.73 V - 0.78 V per cell and a corresponding conversion efficiency of 59-63 \%.  Maintaining the voltage below 0.8 V and keeping it constant, according to a fuel cell catalyst degradation research presented in \cite{thanga}, enables long fuel cell life.

\subsection{Fuel Cell Battery Hybrid System and Control}
Because the load in the experimental system operates on a duty cycle, there is varying power drawn on the system.  A varying load connected directly to fuel cell would cause voltage oscillations that reduce the life of the fuel cell.  For this reason, the fuel cell is connected in parallel with a battery and an electronically controlled variable resistor.  Nominally, the fuel cell trickle charges the battery at constant power, while the battery handles the high and varying demands of the load.  The variable resistor is used to control the voltage during startup/shutdown and if and when the fuel supply drops to limit damage to the fuel cells.

During startup, the fuel cells are held at open circuit voltage for several minutes, providing enough time for hydrogen buildup from the generator before setting the voltage to the operating voltage.  A similar procedure is performed during shutdown and when there is evidence of fuel starvation, where the voltage is set to the open circuit voltage before powering off the remaining components.

A control system is implemented on a micro-controller, an Adruino Mega  (Figure~\ref{fig:fuel_cell_benchtop}).  The voltage of each cell is held at between 0.73 and 0.78 volts.  This is a trade-off to maximize life, power output and fuel cell efficiency. In the next section we analyze the operation of the experiment and demonstrate the feasibility of the concept.

\subsection{Experiment Results and Discussion}

We developed an experimental system to demonstrate the hybrid fuel cell power supply.  The purpose of the experiment is to demonstrate a bench top experiment fuel cell power supply with a fuel cell control system connected to a hydrogen generator that maintains the fuel cell at conditions required to maximize life while producing the power required for a sensor payload. Our objective is to observe whether the fuel cell control system in this hybrid configuration maintains the fuel cell at a desired constant voltage for long durations.  The experiment was run for 5,000 hours, a significant extension over the 1,400 hours demonstrated during our first series of experiments.

\begin{figure} [h!]
\centering
\includegraphics[width=3.5in]{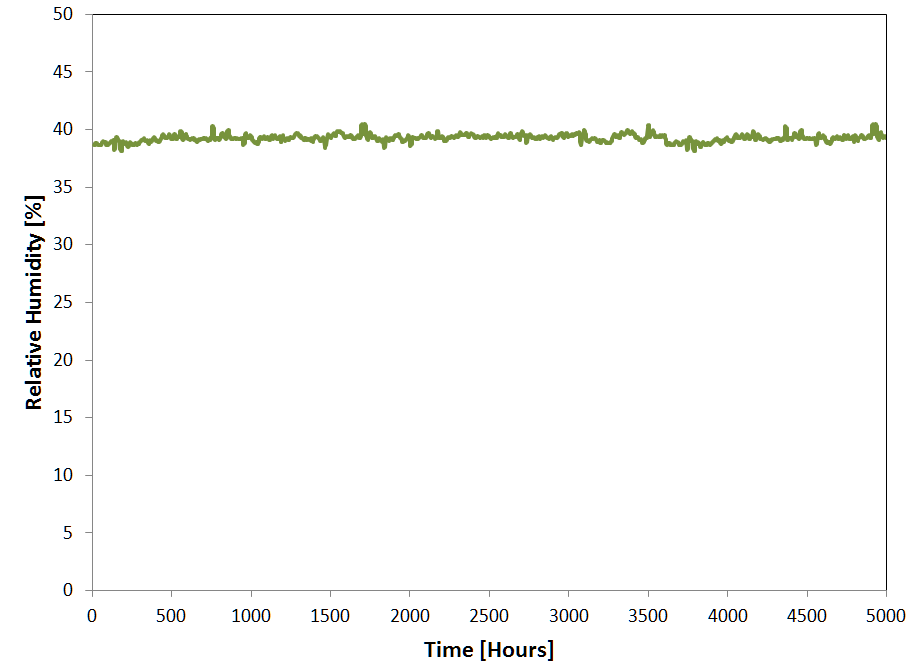}
\caption{Relative humidity experienced by fuel cell power supply bench top experiment.}
 \label{fig:fuel_cell_humidity}
\end{figure}

\begin{figure} [h!]
\centering
\includegraphics[width=3.5in]{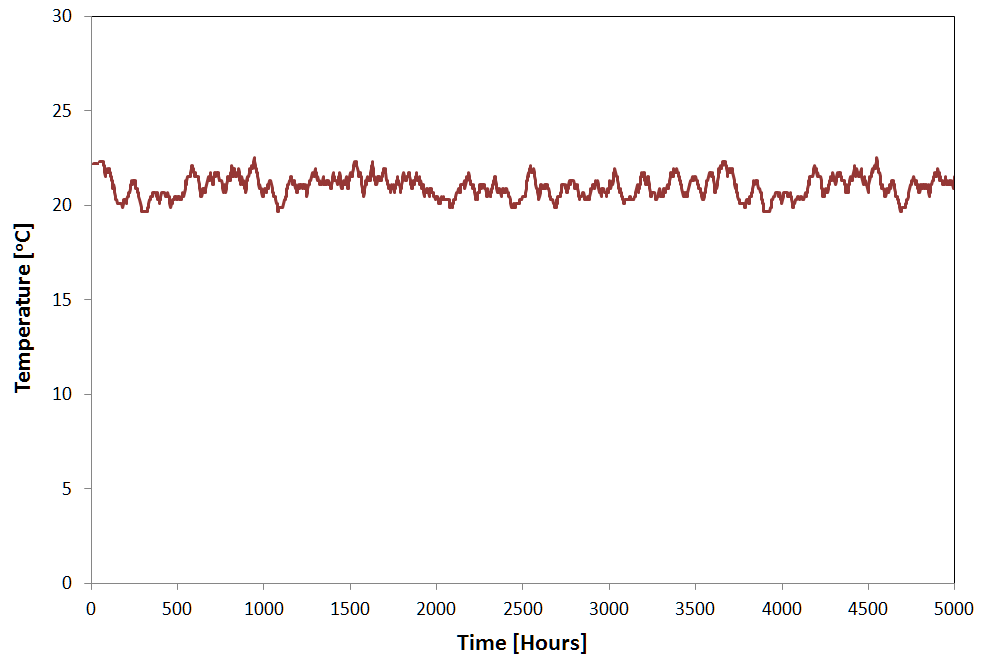}
\caption{Ambient temperature experienced by fuel cell power supply bench top experiment. }
 \label{fig:fuel_cell_temperature}
\end{figure}

\begin{figure} [h!]
\centering
\includegraphics[width=3.5in]{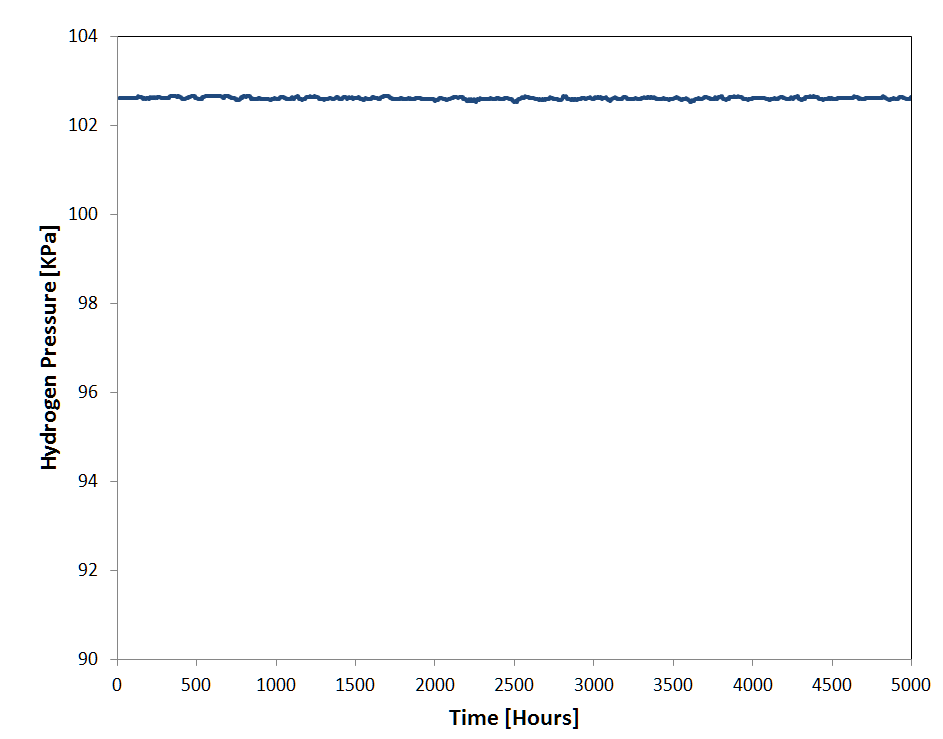}
\caption{Hydrogen pressure from the lithium hydride hydrogen generator.}
 \label{fig:fuel_cell_hydrogen}
\end{figure}

The experimental system recorded several important environmental variables including the environment humidity, ambient temperature, hydrogen pressure, operating voltage and hydrogen pressure for the experimental system.  The relative humidity was controlled and is shown in Figure~\ref{fig:fuel_cell_humidity}.  Figure~\ref{fig:fuel_cell_temperature} shows the ambient temperature.  Figure~\ref{fig:fuel_cell_hydrogen}  shows the hydrogen pressure generated by the hydrogen generator.

\begin{figure} [h!]
\centering
\includegraphics[width=3.5in]{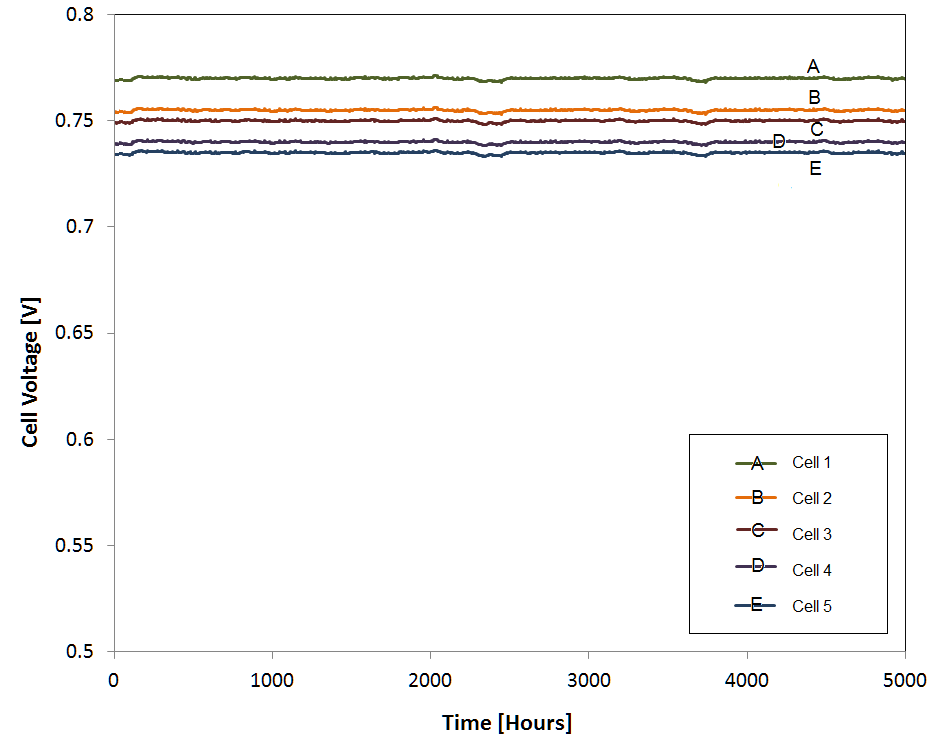}
\caption{Voltage of each cell from the bench top experiment.}
 \label{fig:cell_voltage}
\end{figure}

A control system is implemented on a micro-controller, an Adruino Mega  (Figure~\ref{fig:fuel_cell_benchtop}).  The voltage of each cell is held at between 0.73 and 0.78 volts (Figure~\ref{fig:cell_voltage}).  It should be noted that since the fuel cells are operating in series, the cell nearest to the hydrogen sources has a slightly higher voltage than the cell at the end due to availability of hydrogen.  While the situation is not ideal, it shows that each cell can be maintained with the desired voltage range close to a setpoint.  The voltages are well below 0.8 V and hence minimizes degradation concerns.  A series of trade-off have been made to maximize life, power output and fuel cell efficiency.  With the 5 fuel cell in series, the total voltage ranges from 3.75 to 3.8 Volts.  In turn, the 5 fuel cells are connected in parallel to the battery.  The voltage of the lithium ion battery is a function of how much charge is stored.  Nominally, the sensor load is powered at a nominal $10 \%$ duty cycle, however, if the voltage drops below the threshold 3.70 V or the voltage drops by 0.025 volts per hour, then the controller switches off the load and lets the fuel cell charge the battery to 3.8 V.

If the voltage continues to exceed 3.8 V, then the duty cycle of the sensor load is increased until the equilibrium 3.75 to 3.8 V is reached.  In our experiment, the control system maintains the voltage between 3.75 and 3.8 V (see Figure~\ref{fig:fuel_cell_voltage}).  This is achieved despite the the hydrogen generator operating nominally and provide a constant pressure of hydrogen.  It shown that the fuel cell control system  maintains the fuel cell at the desired operating voltage at all times.

\begin{figure} [h!]
\centering
\includegraphics[width=3.5in]{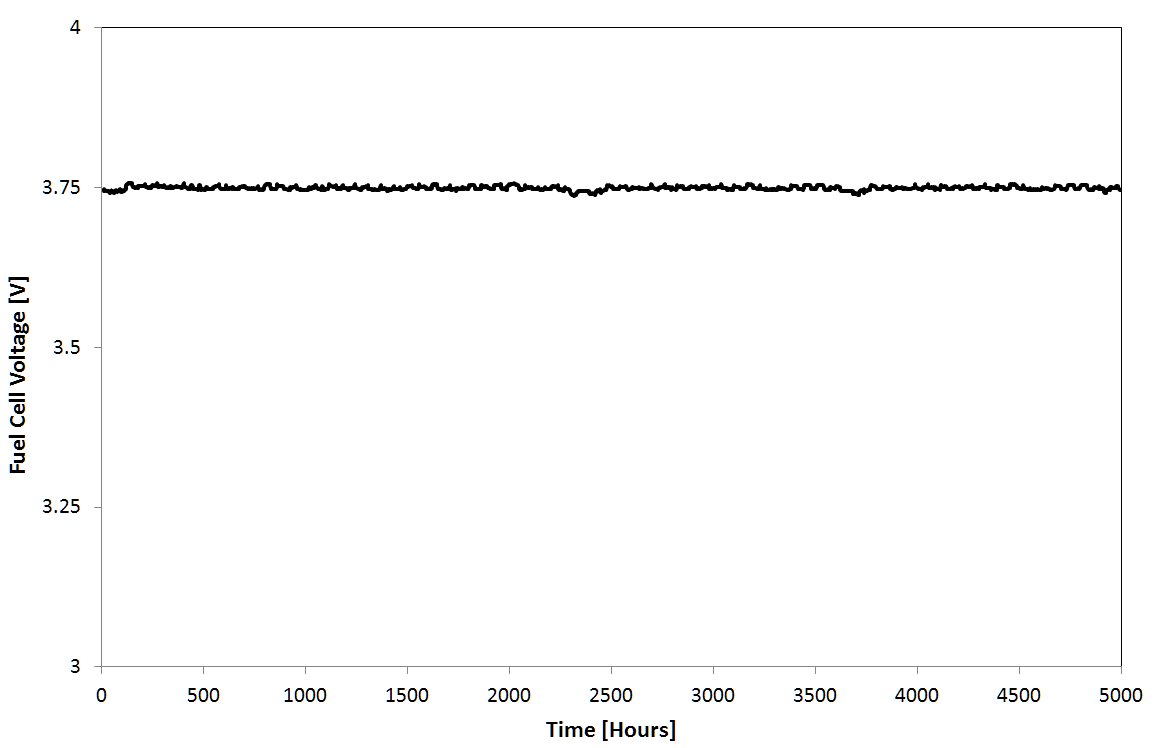}
\caption{Fuel Cell Voltage (cell) from bench top experiment.}
 \label{fig:fuel_cell_voltage}
\end{figure}

The power control system worked as designed maintaining the fuel cell at a constant operating voltage.   This experiment demonstrates that using a simple control algorithm, we can maintain a fuel cell in a hybrid configuration at a nearly constant voltage and have it connected to a hydrogen generator that produces varying amounts of power due to environmental conditions.

After the experiment, the fuel cells were individually tested to determine their performance and if they were impacted by degradation.  Each fuel cell was able to produce 250 mW maximum according to manufacturing specs, which indicated they were unaffected by the experiment.  In other words, there is negligible degradation.  This agrees with our of models from~\cite{thanga}, which shows, for the given operating conditions, we should expect less 5\% degradation after 5000 hours.  The results show that the fuel cell power control system using the battery hybrid configuration is effective at holding a constant voltage and minimizing degradation, provided the necessary hydrogen is supplied.

\section{Case Studies}

\begin{figure*} [t!]
\centering
\includegraphics[width=6.5in]{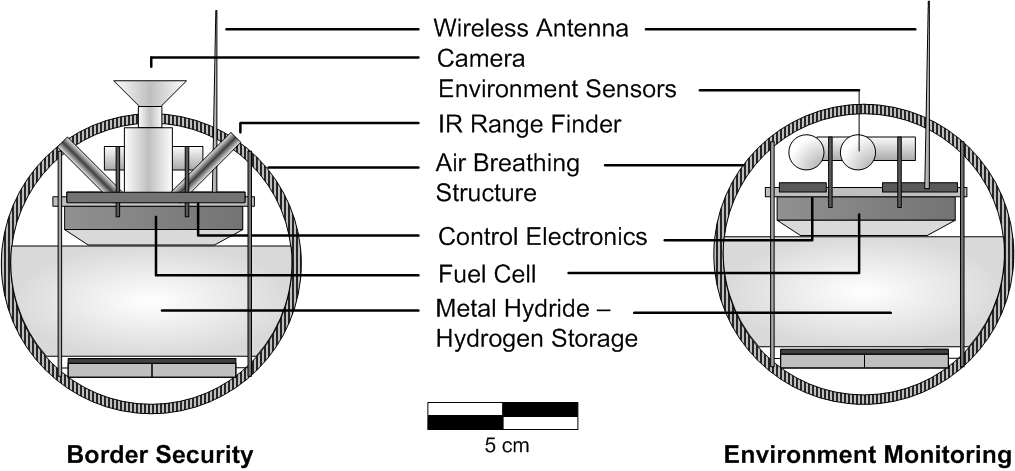}
\caption{Layout of field sensor nodes for application in border security (left) and environmental monitoring (right).}
 \label{fig:sensor_node_layout}
\end{figure*}

Consider a ball shaped sensor network node (Figure~\ref{fig:sensor_node_layout}) with  10-20 cm radius.  The node consists of 4 interchangeable modules, a central CPU module containing electronics and wireless-communication,  a power module consisting of either fuel cells or batteries and a payload modules to house sensors. The payload module may contain temperature, humidity/moisture, vibration, accelerometers, chemical, light sensors and cameras.   These nodes periodically communicate to neighboring nodes and a central base station.  Hence, they will be low power devices that intermittently operate at high power to operate their payloads or communicate data.  Each node will require a minimum 10 mW for standby power.  It is assumed the sensor modules consumes 500 mW average.  These nodes ideally need to operate for 3-5 years unattended, without any periodic maintenance and have a mass less than 30 kg, so that it maybe easily carried and deployed on site.

It is assumed that the field sensor is deployed in a temperate, desert or tropical location, operating continuously, where temperature varies between 15 and 40 $^{o}$ C and humidity varies between 0.15 and 1.0.  In colder climates, the sensor node will require heaters to maintain the fuel cell temperature at 15 $^{o}$ C or higher.

Figure~\ref{fig:sensor_node_layout} shows the sensor node assembled in two different configurations for two different applications.  These applications include (1) Environmental monitoring of air quality and pollution readings (2)  Border security for illegal crossing and smuggling.
For these two different applications, there are different payload sensors.  The key difference between the two application is the operational duty cycle.  For environmental monitoring, the duty cycle can be low, utilizing environmental sensors (air, CO$_{2}$, temperature, humidity, soil moisture) that are typically low power devices with data being gathered periodically at duty cycles of 0.1 or less.  For border monitoring and security applications, the power supply needs to be  fully operational and the sensors need to be constantly operating, thus requiring a high duty cycle.  These sensors may have an infrared range finder, ultrasonic range finder and cameras.  It is presumed state of the art field system is developed consuming an average power of 0.5 W.  However, the actual peak power consumed maybe higher and dependent on the choice of sensors.

\subsection{Battery}

First batteries are considered as power supplies for these nodes. Batteries exhibit self-discharge where stored energy is lost at a fixed rate modeled as a geometric series.  Also, it is assumed that the last 20 \% of the energy cannot be used.  The mass of a battery power supply required is:

\begin{equation}\label{eq:battery}
{M_{bat}} = \frac{{\alpha E(T)\left( {1 - {r^T}} \right)}}{{\rho _{bat}}\left({1 - r}\right)}
\end{equation}

where $M_{bat}$ is the total mass of the battery power supply required for $T$ years of life, $\alpha$ is the capacity margin, $r$ is the self-discharge rate, $\rho_{bat}$ is the energy density of the battery, $E(T)$ is the energy required to power a payload device  for $T$ years according to a given duty cycle. The energy densities, self-discharge rates and mass of the battery power supplies are shown in Table~\ref{tab:bat_tech}.

A sensor module weighing more than 30 kilograms or more lacks scalability to hundreds or thousands of modules owing to high cost and logistics required for deploying/moving them and installing them.  Ideally, the sensor module and power supply needs to have low mass to enable them to be carried in a backpack and deployed in off-grid environments.

\begin{table*}[t]
\centering
\begin{center}
\begin{tabular}{ |l|c|c| }
  \hline
  Technology & Specific Energy (Wh/kg) & Self-Discharge/Degradation (\% per month) \\
  \hline
  Alkaline & 110 & 0.5 \\
  Lithium Ion & 140 & 5 \\
  Lithium CR & 270 & 0.17 \\
  Lithium Thionyl Chloride & 420 & 0.08 \\
  LiH Fuel Cell & 5000 & 0.12 \\
  \hline
\end{tabular}
  \caption{Power Supply Technology Characteristics}
  \label{tab:bat_tech}
\end{center}
\end{table*}

\subsection{PEM Fuel Cell}

Next, the proposed PEM fuel cell power supply concept is compared against batteries. The mass of the fuel for the PEM fuel cell power supply is given by:

\begin{equation}\label{eq:fc}
{M_{fuel}} = \frac{{E(T)}}{{\rho _{fuel}} \cdot r} \cdot \ln \left| {\frac{{0.5 - {\textstyle{1 \over r}}}}{{T + 0.5 - {\textstyle{1 \over r}}}}} \right|
\end{equation}

where $M_{fuel}$ is the total mass of the fuel for $T$ years of life, $\rho_{fuel}$ is the specific energy of the fuel, $E(T)$ is the energy required to power a payload device for $T$ years for a given a power profile, $r$ is the power degradation rate of the fuel cell power supply for a specified operating point.  In addition, the dry mass of the power supply excluding the structural shell is given in Table~\ref{tab:fc_breakdown}.  The lithium hydride fuel has a volume of 0.7g/cm$^3$.  Based on this, the shell consists of two aluminum spheres, 1 mm and 1.5 mm thick, each with enough internal volume to hold the lithium hydride fuel.

\begin{table}
\centering
\begin{center}
\begin{tabular}{ |l|l| }
  \hline
  Component & Mass \\
  \hline
  Fuel Cell and Electronics & 100g \\
  Sensor Payload & 80g \\
  Computer & 50g \\
  \hline
\end{tabular}
  \caption{Dry Mass Breakdown for Fuel Cell Powered Sensor Node}
  \label{tab:fc_breakdown}
\end{center}
\end{table}

The energy density of the fuel is given as follows:
\begin{equation}
{\rho _{Fuel}} = {\rho _{LiH}}\cdot {\lambda _{FC\_EH2}}\cdot {\eta _{LiH\_RC}}
\end{equation}

$\rho_{LiH}$ is the usable quantity of hydrogen energy released from the lithium hydride hydrolysis reaction (presuming water reuse),  $\eta_{LiH\_RC}$ is the percentage reaction completion of the lithium hydride reaction and $\lambda _{FC\_EH2}$ is the efficiency of the fuel cell system.

The total efficiency of the fuel cell is calculated from the following:
\begin{equation}
{\lambda _{FC\_EH2}} = {\lambda _{FC}} \cdot {\lambda _{FC\_Stack}} \cdot {\lambda _{Purge}}
\end{equation}

where $\lambda_{FC}$ is the chemical to electrical efficiency of the individual fuel cells and is related to the operating voltage of the cell, $V/V_{LHV}$, where $V_{LHV}$ is 1.23 V,  $\lambda_{FC_stack}$ is the fuel cell stack efficiency and is 0.95, $\lambda_{purge}$ is the losses due to nitrogen purging and is 0.95.  Our work shows operating each cell at 0.78 Volts, giving it a $\lambda_{FC}$ = 0.63 is a good tradeoff between operating efficiency, fuel cell life and power output.  To supply peak system power and avoid oscillation voltage seen by the fuel cell, the battery handles the high and varying power of the load.  Five micro-fuel cells are required to generate the power required.  Furthermore, the 5 micro-fuel cells configured in series  vastly simplify the fuel cell control electronics and avoids a DC-DC convertor.  With each fuel cell operating at 0.78 volts, the 5 cells are assembled in series to obtain a nominal potential of 3.9 Volts.  Note that the mass calculated accounts for the extra fuel required due to losses from degradation and for ensuring the fuel cell provides the energy required at the end of $T$ years.  Table~\ref{tab:fc_breakdown} shows the mass breakdown of the fuel cell power supply.  The power supply consists of the fuel cell, lithium hydride fuel storage, power control electronics and other components for air and water management.  The lithium hydride fuel produces hydrogen with the addition of water extracted from air.

\subsection{System Comparison between Batteries and Fuel Cells}

\begin{figure} [h!]
\centering
\includegraphics[width=3.5in]{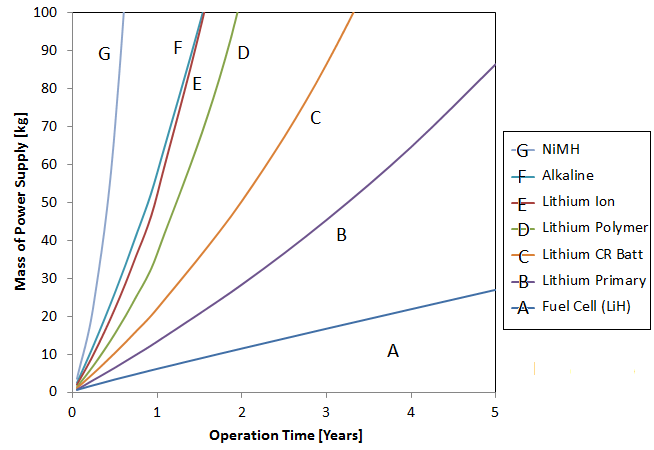}
\caption{Comparison of field sensor power supply technologies for up to 5 years of operation, 0.5 W and 100\% duty cycle.}
 \label{fig:fuel_cell_batt}
\end{figure}

We first compare the proposed system with batteries.  Figure~\ref{fig:fuel_cell_batt} shows the mass of the power system vs average power of 0.5 W for 5 year of  operation at 100\% duty cycle. For very low power, batteries provide an advantage, because of the the additional overhead mass required for fuel cell power supply.  The advantage for the fuel power supply is apparent when the system requires high energy.  Fuel cell using lithium hydride shows a 50 fold advantage in terms of mass compared to lithium ion batteries, a 7 fold advantage over Lithium CR batteries and a 3 fold advantage versus lithium thionyl chloride batteries. In comparison, batteries weigh tens or hundreds of kilograms.  For a network of hundred or thousand nodes, batteries  are not feasible.

\begin{figure} [h!]
\centering
\includegraphics[width=3.5in]{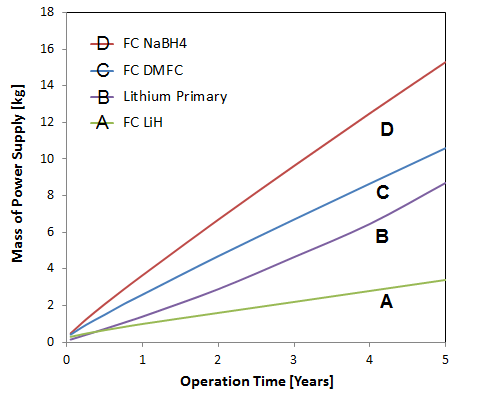}
\caption{Comparison of fuel cell and battery technologies for field sensor power supply.  The system is compared for upto 5 years of operation, 0.5 W and 10\% duty cycle.}
 \label{fig:fuel_cell_low}
\end{figure}

We also compare the  proposed fuel cell power supply with previously reported fuel cell storage technologies (Figure~\ref{fig:fuel_cell_low}).  These previously reported numbers are extrapolated to the required energy for mission lifetime. This includes a PEM fuel cell powered using sodium borohydride and Direct Methanol Fuel Cells. For these comparison, the dry mass for these fuel cell configurations is assumed to be the same as the concept fuel cell system presented here.  In addition, the operating efficiency of direct methanol fuel cell is lower at 40\% and it outputs carbon dioxide that needs to be vented.  DMFC offers a simpler approach to fuel storage, however, challenges exist with fuel cell life, due to buildup of carbon monoxide and low operating efficiencies.  We presume these challenges have been overcome.  The major difference is the mass and volume of the fuel and container.

\begin{figure} [h!]
\centering
\includegraphics[width=3.5in]{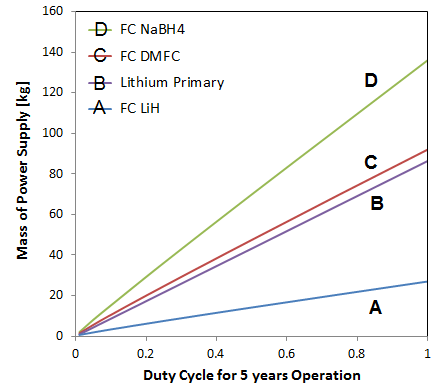}
\caption{Comparison of fuel cell and battery technologies for field sensor power supply.  The system is compared for  5 years of operation for varying duty cycle}
 \label{fig:fuel_cell_lowduty}
\end{figure}

Overall, for low-duty cycle applications (see Figure~\ref{fig:fuel_cell_lowduty}), the proposed fuel cell power supply offers a compelling advantage.  A system of 3.4 kg mass can be fully operational for 5 years at 10\% duty cycle. In comparison, a lithium thionyl chloride primary battery system would weigh 8.7 kg and other fuel cell power supply options weigh  10-15 kg.  Overall, the proposed power supply offers a 2.5 fold mass advantage over the best battery technology and 3 to 4.5 fold advantage over other fuel cell technologies.

However, the required overhead mass for these fuel cell systems is the same. The advantage of the lithium hydride fuel cell
system is reduced mass for long duration missions. Overall, the presented lithium hydride fueled PEM fuel cell offers a substantial advantage for conventional batteries and other fuel cell and hydrogen technologies. These results present a promising pathway towards field testing and evaluation of the proposed concept. For higher duty cycle applications, the proposed system shows an increased advantage over conventional technology.  The mass advantage approaches nearly 3 folds over lithium primary batteries and 3 to 5 folds over other fuel cell technologies.  This shows the promise in this technology for high-energy off-grid applications.

\section{Conclusions}

Field sensor networks have important application in environmental monitoring, border security and infrastructure monitoring.  Current sensor networks rely on solar power  augmented by rechargeable batteries or other methods to harvest energy from the environment.  These systems are bulky and due to varying solar insolation, these sensor networks are not always powered.  Fuel cell power supplies offer a compelling alternative.  They can keep a sensor network fully powered for years at a time, utilizing very little fuel and are clean, quiet and highly efficient.  In this paper, we first analyze how to design a fuel cell to achieve long-life by minimizing the effects of catalyst degradation, while maximizing operating efficiency and performance. Using these techniques, we identified operating conditions that will enable fuel cells to achieve more than 5 years of life, by performing temperature, humidity, operating voltage control and by using a hybrid system that maintains the fuel cells at constant voltage.  The Lithium Hydride fuel can achieve between 4,200 to 4,900 Wh/kg.  Next, we designed and built a bench-top laboratory benchtop system to demonstrate the concept and the system ran for 5,000 hours.  The feasibility of applying the required control techniques were shown under  laboratory conditions.  Based on these laboratory results, we then extrapolate the potential for fuel cell power supplies for various field applications and compare them against current fuel cell and battery technology.  The results show that the proposed technology requires substantially less mass than current battery and conventional fuel cell technologies.  Depending on the applications, these proposed sensor modules can weigh a few kilos, provide up to 0.5 W power and can operate at low duty cycles for 5 years unattended.

\section{Acknowledgements}

The support for this work by Israel's MAFAT Basic Science Office of the MOD is gratefully acknowledged as is the contributions
of Igal Klein, Alex Schechter, Paolo Iora, Kavya Kamal Manyapu, Daniele Gallardo and Ling Ling Deng.

\bibliographystyle{IEEEtran}    
\bibliography{icra_paper}







\end{document}